\documentclass[aps,prb,onecolumn,showpacs]{revtex4}
\usepackage{epsfig}
\usepackage{hyperref}

\begin{document}
\draft 
\def\il{I_{low}} 
\def\iu{I_{up}} 
\def\eeq{\end{equation}}
\def\ie{i.e.}  
\def\etal{{\it et al. }}  
\def\prb{Phys. Rev. {B }}
\def\epjb{Eur. Phys. J. {B }}
\def\pra{Phys. Rev. {A}} 
\def\prl{Phys. Rev. Lett. }
\def\pla{Phys. Lett. A } 
\def\pb{Physica B}
\def\pt{Physics Today }
\def\ajp{Am. J. Phys. }  
\def\mpl{Mod. Phys. Lett. { B}} 
\def\ijmp{Int. J. Mod. Phys. { B}} 
\def\ijp{Ind. J. Phys. }
\def\ijpap{Ind. J. Pure Appl. Phys. }
\def\ibmjrd{IBM J. Res. Dev. }
\def\pjp{Pramana J. Phys.}
\def\ltp{J. of Low Temp. Phys.}
\def\jpcm{J. of Phys. Condensed Matter}

\title{Quantum spin pumping with adiabatically modulated magnetic barrier's}
\author{Ronald Benjamin}
\email{ronald@iopb.res.in}
\affiliation{Institute of Physics, Sachivalaya Marg, Bhubaneswar 751 005,
  Orissa, India}
\author{Colin Benjamin}
\email{colin@iopb.res.in}
\affiliation{Institute of Physics, Sachivalaya Marg, Bhubaneswar 751 005,
  Orissa, India}


\begin{abstract}
  A quantum pump device involving magnetic barriers produced by the
  deposition of ferro magnetic stripes on hetero-structures is
  investigated. The device for dc- transport does not provide
  spin-polarized currents, but in the adiabatic regime, when one
  modulates two independent parameters of this device, spin-up and
  spin-down electrons are driven in opposite directions, with the net
  result being that a finite net spin current is transported with
  negligible charge current. We also analyze our proposed device for
  inelastic-scattering and spin-orbit scattering. Strong spin-orbit
  scattering and more so inelastic scattering have a somewhat
  detrimental effect on spin/charge ratio especially in the strong
  pumping regime. Further we show our pump to be almost noiseless,
  implying an optimal quantum spin pump.

\end{abstract}


\maketitle 

\section {Introduction}

Present day improvements in technology are governed by two major
constraints speed and size. Circuit components are slowly shrinking
while their speed continues to increase. However there is limit to
miniaturization. Making smaller components is not only costly but also
the procedure inherently difficult. Further future miniature devices
are proposed to be built at mesoscopic lengths where unlike recent
times, quantum interference effects will play a major role. A peculiar
and exciting mesoscopic device is the quantum
pump\cite{thou,q_pump_intro,switkes_science} which has been shown to
be adept at implementing rectification\cite{brouwer_rect} and
spin-polarization\cite{spin_wu_wang,taddei}.  Recently a
spin-polarized pump\cite{spin_watson} has also been experimentally
realized based on the theoretical formulations of
Ref. \onlinecite{spin_marcus}. Among the many mesoscopic devices
proposed, those which are effective in providing spin polarized
transport are the most prized, as these are much more resilient to the
vagaries of dephasing.  In this work we propose a quantum spin pump,
aided by the adiabatic modulation of magnetic barriers. A single
magnetic barrier does not provide for spin polarized transport, but
supplemented by adiabatic modulations we can convert it to a cent
percent polarizer.

\begin{figure}[h]
  \protect\centerline{\epsfxsize=3.2in\epsfbox{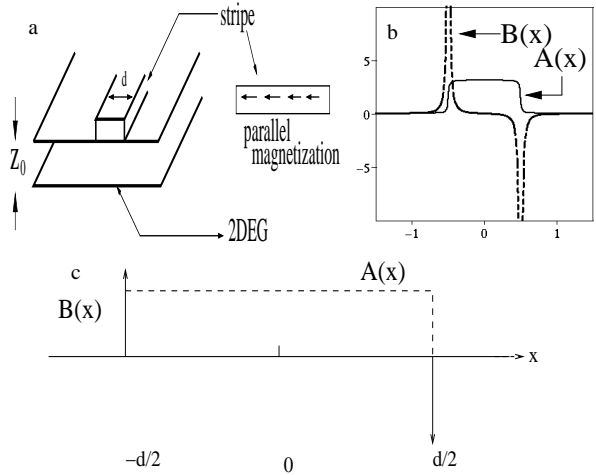}}
\caption{(a) The device- On top of a 2DEG a parallely magnetized
  magnetic stripe is placed. (b) The realistic magnetic field profile
  in a 2DEG along-with the magnetic vector potential for the device
  represented in (a). (c) The model magnetic field (delta function
  B(x)) profile along with the magnetic vector potential A(x).}
\end{figure}

\section {Model}

In this work we propose a spin polarizer based on quantum pumping.
The model of our proposed device is exhibited in Fig.~1. It is
essentially a 2DEG in the $xy$ plane with a magnetic field in the
z-direction. The magnetic field profile we consider is of delta
function type for simplicity, ${\bf B}=B_{z}(x)\hat z$ with $B_{z}(x) =
B_{0}[\delta(x+d/2)-\delta(x-d/2)]$, wherein $B_0$ gives the strength
of the magnetic field and $d$ is the separation between the two
$\delta$ functions (see Fig.1(c)).  The above form of the magnetic
field is an approximation of the more general form seen when parallely
magnetized ferromagnetic materials are lithographically patterned on a
2DEG (Fig.1(b)). Magnetic barrier's can not only be formed by this
method but also when a conduction stripe with current driven through
it is deposited on a 2DEG, and also when a super-conductor plate is
deposited on a 2DEG, see Refs.[\onlinecite{mag_bar_peeter,mag_bar_lu}]
for details.

A 2DEG in the $xy$ plane with a magnetic field pointing in the $z$
direction is described by the Hamiltonian-

\begin{eqnarray}
H&=&\frac{1}{2m^{*}} [{\bf p} +e{\bf A}(x)]^{2} + \frac{eg^{*}}{2m_{0}}
\frac{\sigma \hbar}{2} B_{z}(x)  \nonumber\\
&=&\frac{1}{2m^{*}}
({p_{x}}^{2}+[p_{y}+eA(x)]^{2})+ \frac{eg^{*}}{2m_{0}}\frac{\sigma
  \hbar}{2} B_{z}(x)
\end{eqnarray}

where $m^{*}$ is the effective mass of the electron, $\bf {p}$ is it's
momentum, $g^{*}$ the effective g-factor and $m_{0}$ is the
free-electron mass in vacuum , $\sigma=+1/-1$ for up/down spin
electrons, and ${\bf A}(x)$, the magnetic vector potential is given in
the landau gauge for the region $-d/2< x < d/2$ and for incoming
electrons from the left by ${\bf A}(x)=B_{0}\hat y$, and for electrons
incoming from the right by ${\bf A}(x)=-B_{0}\hat y$, The magnetic
vector potential is zero otherwise. The last term in Eq.~1 is zero
everywhere except at $x=\pm d/2$. For simplicity we introduce
dimensionless units, the electron cyclotron frequency
$w_{c}=eB_{0}/m^{*}c$, and the magnetic length $l_{B}=\sqrt{\hbar
  c/eB_{0}}, with B_{0} $ being some typical magnetic field. All the
quantities are expressed in dimensionless units: the magnetic field
$B_{z}(x)\rightarrow B_{0} B_{z}(x)$, the magnetic vector potential
${\bf{A}(x)}\rightarrow B_{0}l_{B}\bf{A}(x) $, the coordinate
${\bf{x}}\rightarrow l_{b}x$ and the energy $E\rightarrow \hbar w_{c}
E(=E_{0}E)$.

Since the Hamiltonian as depicted in Eq.~1 is translation-ally
invariant along the y-direction, the total wave-function can be
written as $\Psi(x,y)=e^{i q y} \psi(x)$, wherein $q$ is the
wave-vector component in the y-direction. Thus one obtains the
effective one-dimensional Schroedinger equation-

\begin{equation}
[\frac{d^{2}}{dx^{2}}-\{A(x)+q\}^{2}- \frac{eg^{*}}{2m_{0}}\frac{\sigma
  m^{*}}{\hbar} {B_{z}(x)} + \frac{2 m^{*}}{\hbar^{2}} E] \psi(x)=0 
\end{equation}

\section {Theory}
The S-matrix for electron transport across the device can be readily
found out by matching the wave functions and as there are $\delta$
function potentials there is a discontinuity in the first derivative.
The wave functions on the left and right are given by $\psi_1
=(e^{ik_{1}x}+r e^{-ik_{1}x})$ and $\psi _3=t e^{ik_{1}x}$, while that
in the region $-d/2<x<d/2$ is $\psi_2=(a e^{ik_{2}x}+b e^{-ik_{2}x})$.
The wave vectors are given by- $k_{1}=\sqrt{2E-q^2}$,
$k_{2}=\sqrt{2E-(q+B_{z})^2}$ and for electrons incident from the
right ,$k2$ in the wave-functions is replaced by
$k_{2}^{\prime}=\sqrt{2E-(q-B_{z})^2}$. Throughout this article,
unless specified otherwise, $q=0$, and therefore
$k^{\prime}_{2}=k_{2}$.

With this procedure outlined above one can determine all the
coefficients of the S-Matrix

\[S_{\sigma}=\left(\begin{array}{cc}
s_{\sigma 11}       & s_{\sigma 12} \\
s_{\sigma 21} &     s_{\sigma 22}  \\
\end{array} \right)=\left(\begin{array}{cc}
r_{\sigma}       & t^{\prime}_{\sigma} \\
t_{\sigma} &     r^{\prime}_{\sigma}  \\
\end{array} \right) \]

One can readily see from the transmission coefficients, for details
see Ref.\onlinecite{sm} that there is no spin polarization as
$T_{+1}=T_{-1}$. This fact was discovered only in
Ref.\onlinecite{mag_bar_papp_err}, two earlier
works\cite{mag_bar_amlan,mag_bar_papp} had mistakenly attributed spin
polarizability properties to the device depicted in FIG.~1.  In the
adiabatic regime, the device is in equilibrium, and for it to
transport current one needs to simultaneously vary two system
parameters $X_{1}(t)=X_{1}+\delta X_{1}sin(wt)$ and
$X_{2}(t)=X_{2}+\delta X_{2}sin(wt+\phi)$, in our case $X_1$ is the
width $d$ and $X_2$ the magnetic field $B_z$ given in terms of the
magnetization strength $B_{0}=M_{0}h$, where $h$ is the height and
$M_0$ the magnetization of the ferro-magnetic stripe.

The pumped current can be calculated by using the procedure as adopted
in Ref. \onlinecite{q_pump_brouwer} and in Ref. \onlinecite{q_pump_wang}
for the case of a double barrier quantum well. A new formalism taking
recourse to Floquet theory\cite{flouquet} has recently been applied to
describe quantum pumping but in the following discussion we will
concentrate only on Brouwer's approach as elucidated in
Ref. \onlinecite{q_pump_brouwer}. This approach has been further
applied to several different systems, among them mention may be made
of- quantum pumping in carbon nanotubes\cite{wang_nano}, quantum
pumping in systems with a super-conducting lead attached\cite{blau}and
study of dephasing in quantum pumps\cite{cremers,mosk}.

In the succeeding discussion, unless specified otherwise $\alpha=1$,
i.e., we always pumping into the left lead or channel $1$, (left of
the barrier at $-d/2$, see Fig.~1(c) ). The right lead or channel $2$ is
to the right of the barrier at $d/2$. We further assume single moded
transport in the leads or channels.  Thus charge passing through lead
$\alpha$ due to infinitesimal change of system parameters is given by-

\begin{eqnarray}
dQ_{\sigma \alpha}(t)=e [\frac{dN_{\sigma \alpha}}{dX_1} \delta X_{1}(t)+\frac{dN_{\sigma \alpha}}{dX_2} \delta X_{2}(t)]
\end{eqnarray}

with the current transported in one period being-

\begin{eqnarray}
I_{\sigma \alpha}=\frac{ew}{2\pi}\int_{0}^{\tau} dt [\frac{dN_{\sigma \alpha}}{dX_1}\frac{dX_{1}}{dt}+
\frac{dN_{\sigma \alpha}}{dX_2}\frac{dX_{2}}{dt}]
\end{eqnarray}

In the above $\tau=2\pi/w$ is the cyclic period. The quantity
$dN_{\sigma\alpha}/dX_i$ is the emissivity which is determined from
the elements of the scattering matrix, in the zero temperature limit
by -

\begin{eqnarray}
\frac{dN_{\sigma \alpha}}{dX_i}=\frac{1}{2\pi}\sum_{\beta} \Im (\frac{\partial s^{}_{\sigma \alpha \beta}}{\partial X_i}  s^{*}_{\sigma \alpha \beta})
\end{eqnarray}

Here $s_{\sigma \alpha \beta}$ denote the elements of the scattering
matrix as denoted above, as evident $ \alpha, \beta$ and $i$ can only
take values 1,2, while $\sigma$ takes values $+1$ or $-1$ depending on
whether spin is up or down. The symbol ``$\Im$'' represents the
imaginary part of the complex quantity inside parenthesis.

The spin pump we consider is operated by changing the width and
magnetic field strength $B_z$ (given in terms of magnetization
$B_0=M_{0}h$) of the ferro magnetic stripe, herein
$X_{1}=d=d_{0}+x_{p}sin(wt)$ and
$X_{2}=B_{z}=B_{x}+x_{p}sin(wt+\phi)$. A paragraph on the experimental
feasibility of the proposed device is given above the conclusion. As
the pumped current is directly proportional to $w$ (the pumping
frequency), we can set it to be equal to $1$ without any loss of
generality.

By using Stoke's theorem on a two dimensional plane, one can change
the line integral of Eq.~4 into an area integral, see for details
Ref.\onlinecite{cheng}-
\begin{eqnarray}
I_{\sigma \alpha}=e\int_{A} dX_{1} dX_{2} [\frac{\partial}{\partial X_{1}}\frac{dN_{\sigma \alpha}}{dX_2}-
\frac{\partial}{\partial X_{2}}\frac{dN_{\sigma \alpha}}{dX_1}]
\end{eqnarray}

Substitution of Eq.~5 into Eq.~6 leads to,

\begin{eqnarray}
I_{\sigma \alpha}=e\int_{A} dX_{1} dX_{2} \sum_{\beta=1,2} \Im (\frac{\partial s^{*}_{\sigma \alpha \beta}}{\partial X_1} \frac{\partial s^{}_{\sigma \alpha \beta}}{\partial X_2})
\end{eqnarray}

If the amplitude of oscillation is small, i.e., for sufficiently weak
pumping ($\delta X_{i} \ll X_{i}$), we have,

\begin{eqnarray}
I_{\sigma \alpha}=\frac{ew\delta X_{1}\delta X_{2}sin(\phi)}{2\pi} \sum_{\beta=1,2} \Im (\frac{\partial s^{*}_{\sigma \alpha \beta}}{\partial X_1} \frac{\partial s^{}_{\sigma \alpha \beta}}{\partial X_2})
\end{eqnarray}

In the considered case of a magnetic barrier the case of very weak
pumping is defined by: $x_{p} \ll B_{x}(=d_{0})$, and Eq.~8 becomes-

 \begin{eqnarray}
I_{\sigma \alpha}=I_{0} \sum_{\beta=1,2} \Im (\frac{\partial s^{*}_{\sigma \alpha \beta}}{\partial B_{z}} \frac{\partial s^{}_{\sigma \alpha \beta}}{\partial d})
\end{eqnarray}
wherein,
\[I_{0}=\frac{ew x^{2}_{p} sin(\phi)}{2\pi}\] 
As we consider only the pumped currents into lead 1, therefore
$\alpha=1$. Further we drop the $\alpha$ index in expressions below.
From the elements of the S-Matrix given in Ref. \onlinecite{sm}, one
can easily derive analytical expressions for the pumped current
$I_\sigma$, pumped spin $I_{sp}$ and charge $I_{ch}$ currents in the
very weak pumping limit addressed in Eq.~9, as follows-

\begin{eqnarray}
I_{\sigma}&=&\sigma I_{0}\frac{2B^{2}_{z}g^{*}g^{\prime}k^{3}_{1}k^{2}_2sin(2k_{2}d)}{T^{2}_d},\\
I_{sp}=I_{+1}-I_{-1}&=&I_{0}\frac{4B^{2}_{z}g^{*}g^{\prime}k^{3}_{1}k^{2}_2sin(2k_{2}d)}{T^{2}_d},\\
I_{ch}=I_{+1}+I_{-1}&=&0,\\
\mbox{with }g^{\prime}=1-\frac{g^{*2}}{4}, T_{d}&=&4k^{2}_{1}k^{2}_{2}cos^{2}(k_{2}d)+[4E-g^{\prime}B^{2}_{z}]^{2}sin^{2}(k_{2}d),\nonumber\\
\mbox{and the wave-vectors are given by-} \nonumber\\
k_{1}&=&\sqrt{2E} \mbox{ and } k_{2}=\sqrt{2E-B^{2}_{z}}\nonumber
\end{eqnarray}

It should be noted that the pumped charge current is identically zero,
as the terms in the expression for $I_{ch}$ cancel out resulting in
zero pumped charge current in the weak pumping regime. In the
succeeding sections we analyze the pumped spin and charge currents for
different variations of parameters, in both the sufficiently weak
pumping case (Eq.~9), as well as the general case of weak to strong
pumping (Eq.~4).

\begin{figure*}[!]
  \protect\centerline{\epsfxsize=7.0in\epsfbox{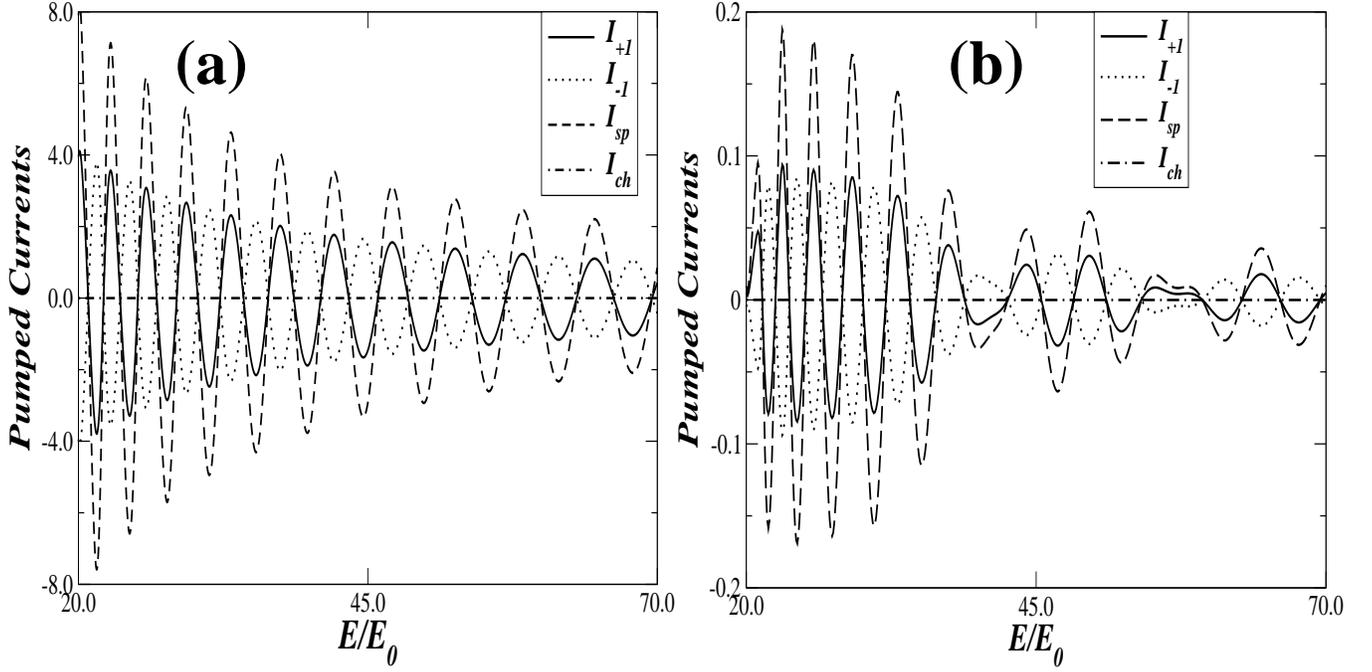}}
\caption{Energy dependence of the pumped current. Spin polarized
  pumping delivering a net spin current along-with a vanishing charge
  current.  The parameters are $B_{x}=5.0, d_{0}=5.0, \phi=\pi/2,
  g^{*}=0.44$ and wave-vector $q=0$. (a) The weak pumping regime. The
  pumped currents are normalized by $I_0$. In (b) the case of strong
  pumping for $x_{p}=1.0$ is plotted.}
\end{figure*}

\section {Characteristics of the Pumped Current}

In Fig.~2, we plot the pumped currents (magnified 100 times in Fig.
2(b)) as function of the Fermi energy at zero temperature for spin-up
$I_{+1}$ (solid-line), spin-down $I_{-1}$ (dotted line),
spin-polarized $I_{sp}=I_{+1}-I_{-1}$ (dashed line) and the net charge
current $I_{ch}=I_{+1}+I_{-1}$ (dot-dashed line) in the special case
of (a) very weak pumping (Eq.~9) and for the general case (Eq.~4) in
(b).  Again, unless specified otherwise, throughout the discussion
temperature is always zero. The net charge current is expressed in
terms of the electric charge. The parameters in dimensionless units
are mentioned in the figure caption. The modulated parameters are out
of phase by $\pi/2$.  From Fig.2(a) and (b) it is evident that
throughout the range of the Fermi-energy the net pumped charge current
is negligible while a non-zero spin current is pumped. Quantitatively,
for the general case of Fig.2(b) and for a $GaAs$ based system,
$g^{*}=0.44, m^{*}=0.067 m_{e}$ and if $B_{0}=$0.1T then $l=813\AA,
\hbar w_{c}=E_{0}=0.17 mev$, thus for parameters in Fig.~2(b), energy
$E\sim 4.0-12.0 mev$, magnetic field strength $B_{x}= 0.5$T,
$x_{p}$=0.1T, and if $w$ of the order of $10^{8}$ Hz (as in the
experimental arrangement of Switkes, et.al. in Ref.
\onlinecite{switkes_science}), then pumped spin current $I_{sp}=1.6
\times 10^{-19}C \times 10^{8} Hz \times 0.005 \sim 1 \times 10^{-13}
Amperes$, while the pumped charge current is $10^4$ times weaker
around $10^{-17} Amperes$. In Fig.~3 we plot the pumped spin and
charge currents (from Eq.~4) as a function of $x_{p}$, the pumping
amplitude for weak to strong pumping. We find a finite spin current
with negligible charge current throughout the range of the pumping
amplitude from the very weak to very strong. The figure 3, also
conveys the very important fact that for the entire range, from the
very weak to the very strong pumping regimes, we see increase in the
magnitude of pumped currents which suggests that our model device
would pump large spin currents in the very strong pumping pumping
regime. Further the pumped charge current is, for throughout the range
of the pumping amplitude, zero. The physics behind the pumping
mechanism in our model is as follows, in case of dc transport the
transmittance is even in spin, as the Hamiltonian (Eq.~1) is time
reversal invariant, as a consequence there is no spin
polarization\cite{dobro}, but herein as we consider the adiabatic
modulation procedure, with the condition that the pumping amplitudes
are out of phase by $\phi$, which implies the dynamical breaking of
time reversal invariance which in turn leads to a net spin current
being pumped. Recently it has been shown that for a ring with an
oscillating scatterer (wherein potentials oscillate out of phase) the
time reversal symmetry is dynamically broken and hence a net
circulating (pumped) current arises\cite{mosk_pump_circ}.
\begin{figure}[!]
\protect\centerline{\epsfxsize=4.0in\epsfbox{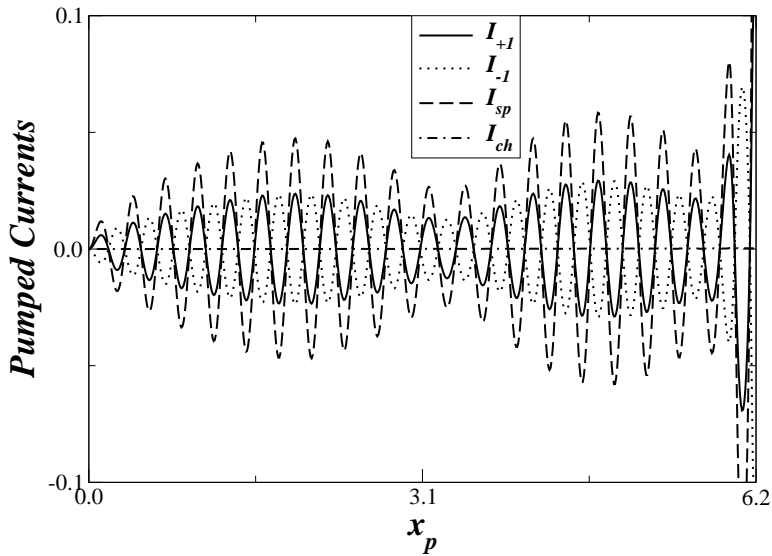}}
\caption{ Dependence of the pumped current on amplitude of pumping. 
  Spin polarized pumping delivering a net spin current along-with a
  vanishing charge current. The parameters are $d_{0}=5.0, B_{x}=5.0,
  E/E_{0}=64.3, \phi=\pi/2, g^{*}=0.44$ and wave-vector $q=0$.}
\end{figure}

The pumped currents are sinusoidal as function of the phase difference
for the weak pumping regime, but for the strong pumping regime this
sinusoidal behavior is absent. In Fig.~4, we plot the pumped currents
as a function of the phase difference $\phi$ for Fermi energy
$E/E_{0}=23.12$, for the case of weak pumping $x_{p}=0.1$ in (a), and
the general case of strong pumping for $x_{p}=1.0$ in (b). In
accordance with the results for a generic double barrier quantum pump,
the pumped currents are anti-symmetric about $\phi=\pi$ and maximum at
$\phi=\pi/2$ in (a), but in (b) we see that the that the relation
between pumped currents and the phase difference is non-sinusoidal
although they are still anti-symmetric about $\phi=\pi$ and the
currents peak at small difference in phase.  In the strong pumping
regime as is wont the magnitude of pumped currents are much larger
than in the weak pumping regime, but in both the strong as well as
weak pumping regimes the pumped currents are periodic with period
$2\pi$. The figure 4 also conveys the important fact that throughout
the range of the phase difference $\phi$ we see zero pumped charge
current, while the pumped spin currents are non-zero, and in figure
4(b) for the case of strong pumping the spin currents are much larger.

\begin{figure*}[!]
\protect\centerline{\epsfxsize=7.0in\epsfbox{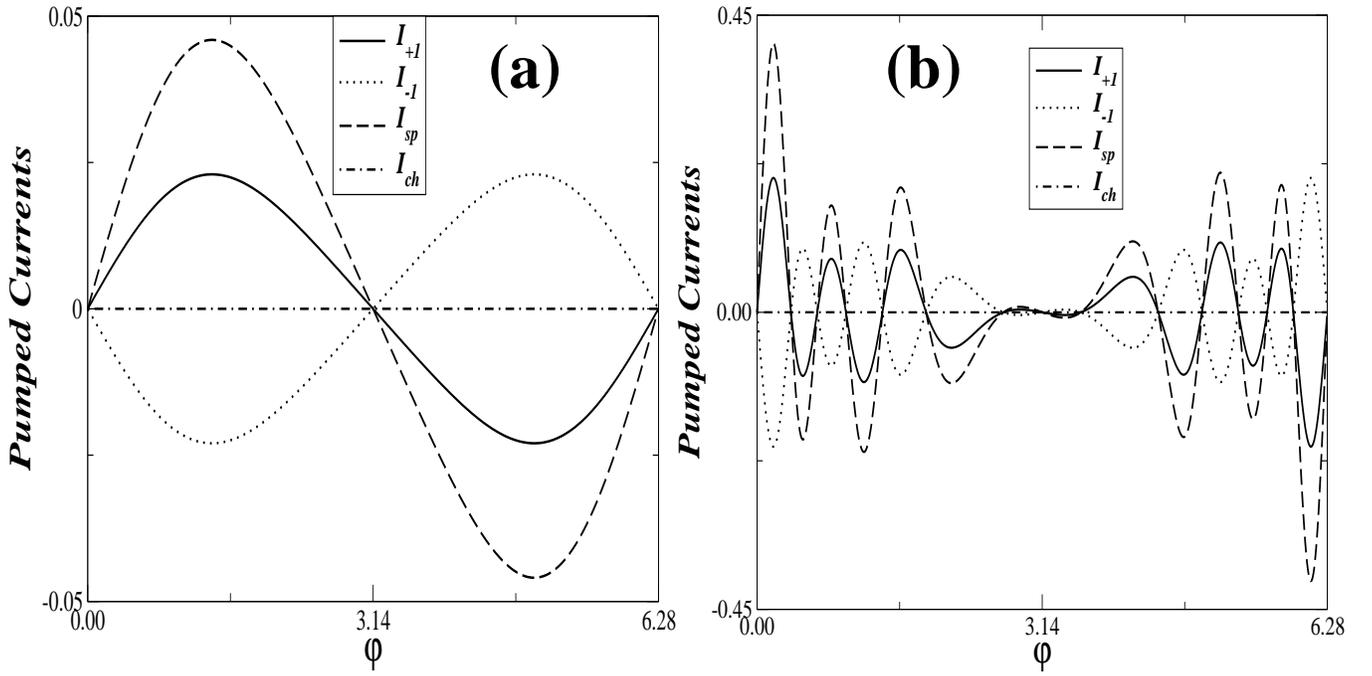}}
\caption{ Dependence of the pumped current on phase difference
  $\phi$.  Spin polarized pumping delivering a net spin current
  along-with a vanishing charge current.(a) Weak pumping regime. The
  parameters are $d_{0}=B_{x}=5.0, x_{p}=.1, E/E_{0}=23.12$ and
  wave-vector $q=0$.(b)Strong pumping regime.  The pumped currents are
  plotted for $x_{p}=1$ all other parameters remaining same.}
\end{figure*}

\begin{figure*}[!]
\protect\centerline{\epsfxsize=7.0in\epsfbox{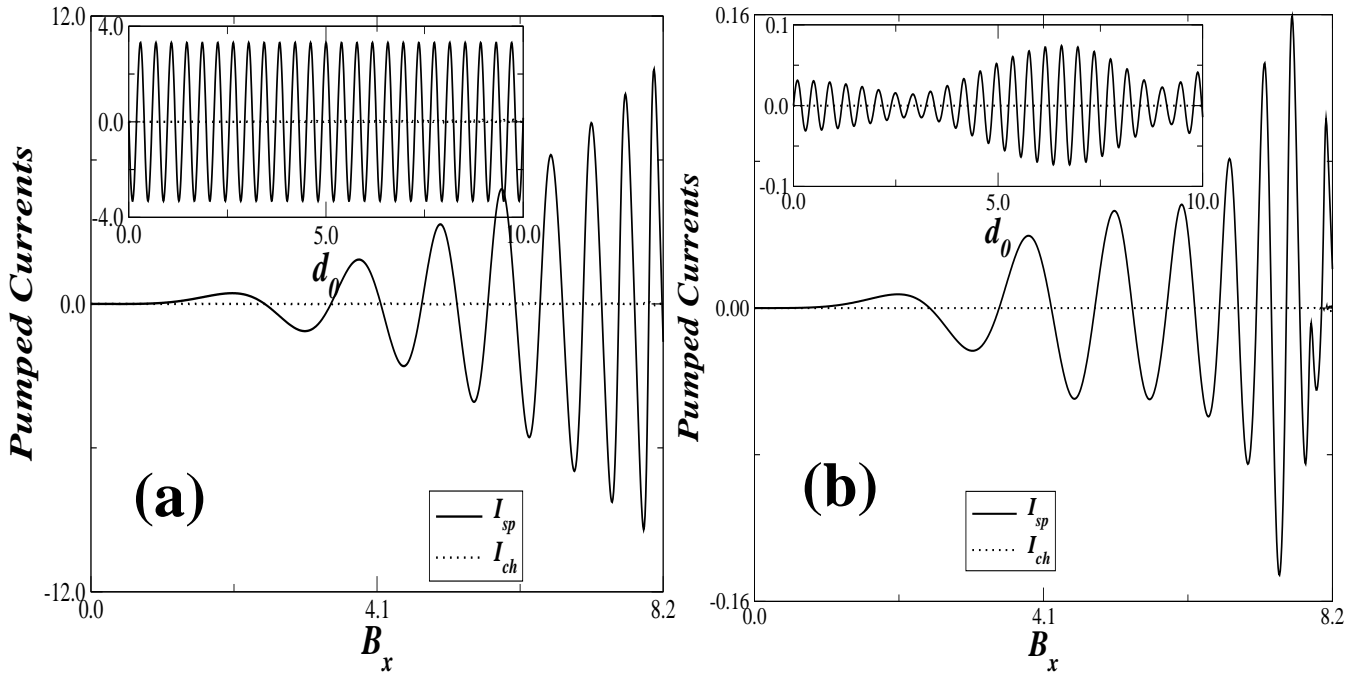}}
\caption{ Dependence of the pumped current on amplitude of barrier strength
  $B_{x}$. Spin polarized pumping delivering a net spin current
  along-with a vanishing charge current.(a)Weak pumping regime.The
  parameters are $d_{0}=5.0, \phi=\pi/2, E/E_{0}=44.6$ and wave-vector
  $q=0$. The pumped currents are normalized by $I_0$. (b) Strong
  Pumping regime $x_{p}=1.0$. In the inset of (a) and (b) the pumped
  currents are plotted as function of the width $d_{0}$, $B_{x}=5.0,$
  all other parameters remaining same.}
\end{figure*}

\section {Importance of Resonances}
Resonances play an important role in the case of quantum pumping as
exemplified in Refs.[\onlinecite{q_pump_wang,wang_reso}], in the
following we depict the variation of the pumped currents with magnetic
barrier strength and width of the magnetic barrier and show that this
is indeed the case here also.  In FIG.~5(a), we plot the pumped
currents as a function of $B_x$, i.e., the strength of the magnetic
barrier, for Fermi energy $E/E_{0}=44.6$ and phase difference
$\phi=\pi/2$ for the special case of very weak pumping as in Eq.~9 and
in Fig.5(b) for the general case as in Eq.~4. The pumped currents
depend on the strength of the barrier and for increased barrier
strength these seem to be larger. As the Fermi energy is set at
$E/E_{0}=44.6$, naturally a magnetic barrier of height of the order
$\sim 5.0$ or more will affect the electron and so naturally one sees
increased pumping for larger values of $B_x$.  In the inset of
FIG.~5(a) and (b), we plot the pumped currents as a function of the
$d_0$, i.e., the width of the magnetic barrier, for Fermi energy
$E/E_{0}=44.6$. The pumped currents have almost a nice sinusoidal
dependence on the width. Herein also as the resonances are controlled
by the width one can explain these sinusoidal variations on the
resonances of the system. From the analytical expression for the
pumped currents (Eqs.~10-12) in the weak pumping regime one can easily
notice that this sinusoidal dependence arises because of the
$sin(2k_{2}d)$ factor in the numerator of Eqs.[10,11]. One can
approximate $g^{\prime} \sim 1$ as $g^{*}=0.44$ for $GaAs$ based
system and thus the pumped current becomes-
\begin{eqnarray}
I_{\sigma}\sim \sigma I_{0}\frac{\sqrt{2E}B^{2}_{z}g^{*}sin(2k_{2}d)}{16E(2E-B^{2}_{z})[1+\frac{B^{4}_{z}}{8E(2E-B^{2}_{z})} sin^{2}(k_{2}d)]^{2}}
\end{eqnarray}

When $2E \gg B^{2}_z$ one can neglect the second term inside the
square bracket in the denominator of Eq. 13 as it is very small. Thus
the pumped current in this limit reduces to
\[I_{\sigma} \sim \sigma I_{0} \frac{B^{2}_{z} g^{*}
  sin(2k_{2}d)}{16E\sqrt{2E}}.\] Thus we get the condition for
resonances as $2k_{2}d=(2n+1)\pi/2, n=0,1,2\ldots$ . The approximate
position of the resonances in the $2E >> B^2_z$ regime, occur at Fermi
energies- $E_{n}=\frac{(2n+1)\pi}{8d} +\frac{B^2_z}{2}, n=0,1,2\ldots
$ .

\begin{figure*}[b]
\protect\centerline{\epsfxsize=7.0in\epsfbox{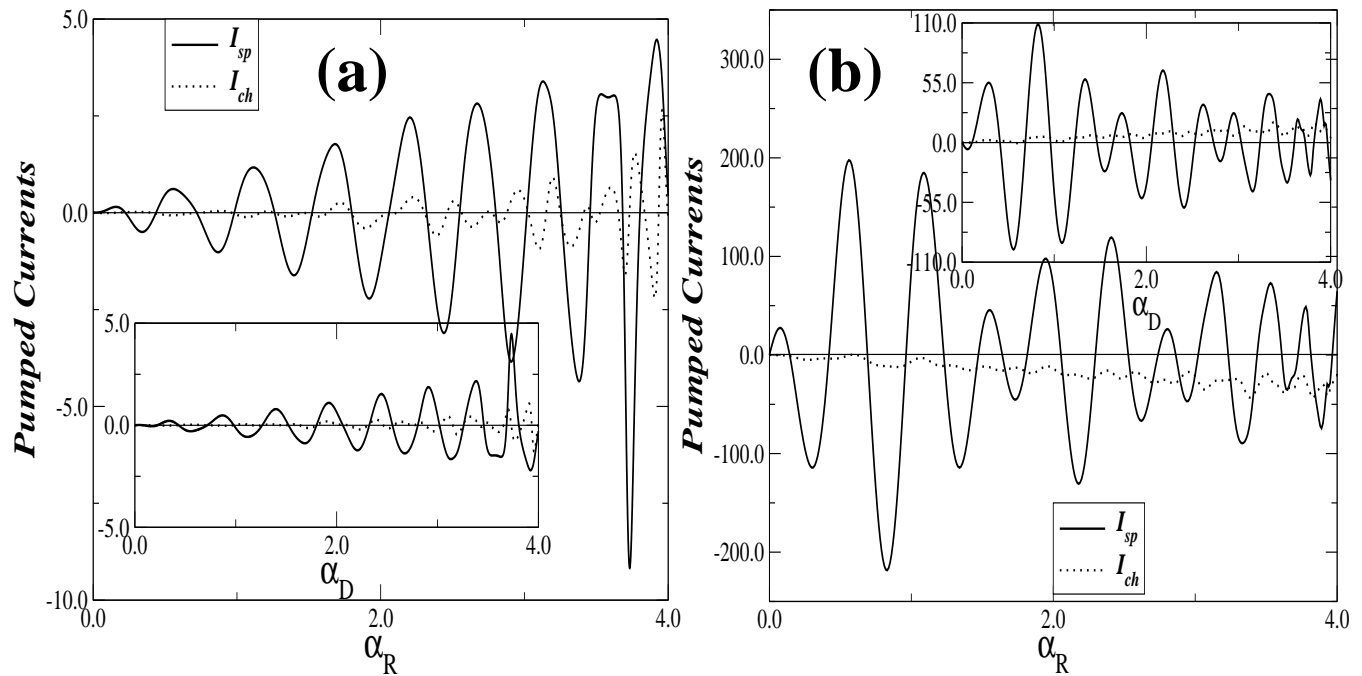}}
\caption{Dependence of the pumped current on Rashba spin-orbit interaction
  $\alpha_R$. Spin polarized pumping delivering a net spin current
  along-with a small finite charge current. (a) Very weak pumping
  regime. The pumped currents are normalized by $I_0$. The parameters
  are $B_{x}=5.0, d_{0}=5.0, E/E_{0}=64.3, \phi=\pi/2$ and wave-vector
  $q=0$. In the inset the dependence of the pumped currents on the
  Dresselhaus spin-orbit interaction $\alpha_D$ is plotted, parameters
  remaining same. (b) Strong pumping case $x_{p}=1.0$ all other
  parameters remaining same. In the inset the dependence of the pumped
  currents on the Dresselhaus spin-orbit interaction $\alpha_D$ is
  plotted, parameters remaining same.}
\end{figure*}

\section {Spin-Orbit Scattering}
In the above analysis we have ignored the effects of spin-orbit
scattering as this is generally supposed to be very small in these
systems. However for a complete theory we have to include the effects
of spin-orbit scattering, and analyze it's impact on spin pumping. The
spin-orbit scattering (SO) can arise due to two reasons\cite{moroz}-
{\it (i) Microscopic forces (Dresselhaus effect)-} In general III-V
compounds (e.g., GaAs) lack inversion symmetry. This eventually leads
to spin-orbit scattering induced splitting of the conduction
band\cite{dress}. The magnitude of the splitting is proportional to
cube of electron wave number $k$. In MOSFET's and hetero-structures,
the host crystals are not treated as 3D systems, because crystal
symmetry is broken at the interface where the 2DEG is dynamically
confined in a quantum well. The reduction of effective dimensionality
lowers symmetry of underlying crystals and results in an additional
(linear in $k$) term in the Dresselhaus splitting. It is seen that the
linear in $k$ term is dominant in GaAs quantum wells.  {\it (ii)
  Macroscopic forces (Rashba effect)-} In addition to the above
microscopic forces there is another source of splitting, an interface
electric field\cite{rashba}. It is manifest in a linear in $k$
splitting of 2D band structure. In most 2DEG systems, the Rashba term
dominates the Dresselhaus terms. Herein we first consider the Rashba
spin-orbit interaction. In presence of this Rashba spin-orbit
interaction the Hamiltonian defined in Eq.~1 is modified with the
addition of the following
term-$H_{R}=\alpha_{R}(\sigma_{y}p_{x}-\sigma_{x}p_{y})$.  With this
addition into the Hamiltonian only in region II (we assume the
spin-orbit interaction only in the confines of the magnetic barrier),
the wave-function is now described by two eigen-vectors corresponding
to the Rashba split eigenvalues
$E_{1(2)}=E\pm\alpha_{R}\sqrt{2E+(q+B_{z})^2}$. Similar to the
previous example one solves for the reflection and transmission
amplitudes (see for details Ref.\onlinecite{jalil}) and calculates the
pumped currents. Results are shown in Fig.~6(a) for the special case
of very weak pumping (from Eq.~9) and in Fig.~6(b) for the general
case (from Eq.~4).  The inclusion of Rashba spin-orbit interaction
leads to no change when $\alpha_{R}$, is small, i.e., the spin-orbit
scattering length ($l_{so}\sim \frac{1}{\alpha_R}$) is large. But
increasing $\alpha_{R}$ leads to a small charge current, which for
$l_{so}\ll d$ becomes significant, and of same order of magnitude as
the spin current. Interestingly the spin-current oscillates as a
function of $\alpha_R$, indicating the importance of interference
effects. In the inset of Fig's.~6(a) and 6(b), we have depicted the
effect of linear in $k$ Dresselhaus type spin-orbit interaction,
$H_{D}=\alpha_{D}(\sigma_{x}p_{x}-\sigma_{y}p_{y})$.  Similar to the
Rashba type in this case also pumped currents behave correspondingly,
as a function of the Dresselhaus spin-orbit interaction strength
$\alpha_{D}$. It would also be worthwhile to point out that in-spite
of the fact that a small charge current contribution manifests itself
in the strong pumping regime, the magnitude of the pumped spin
currents increases manifold both in case of Dresselhaus spin-orbit
interaction and more so in case of Rashba spin-orbit interaction.

\section {Inelastic Scattering}
Inelastic scattering has been ignored in our discussion so far, as we
assume that the electron retains it's phase coherence throughout the
sample. This assumption has of-course limited validity as for low
temperatures electron-phonon scattering is absent but
electron-electron scattering is always present and may lead
randomization of phase implying incoherent scattering and resulting in
loss of coherence. To include inelastic (or incoherent) scattering so
as to see it's effect on the spin/charge ratio we follow the formalism
developed in Ref.\onlinecite{mosk}. In this formalism a third
fictitious voltage probe is coupled to the quantum pump and all
inelastic processes culminating in dephasing are described by a single
parameter $\epsilon$.
\begin{figure}[h]
\protect\centerline{\epsfxsize=4.0in\epsfbox{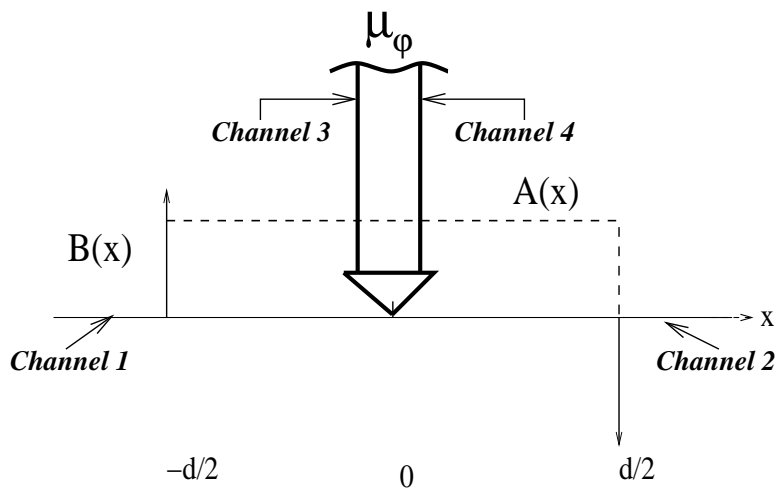}}
\caption{The model magnetic field (delta function B(x))  profile along
  with the magnetic vector potential A(x) in presence of a voltage
  probe$\mu_\phi$ attached to inelastic channels $3$ and $4$. }
\end{figure}
In this model the Eq.~4, is modified to take into account inelastic
processes in the following manner-

\begin{eqnarray}
I_{\sigma 1}=\frac{ew}{2\pi}\int_{0}^{\tau} dt [F_{\sigma 1}+K_{in,
  \sigma 1}(F_{\sigma 3}+F_{\sigma 4})]
\end{eqnarray}

with the charge pumped given by
\[
F_{\sigma \alpha}=\frac{dN_{\sigma \alpha}}{dX_1}\frac{dX_{1}}{dt}+
\frac{dN_{\sigma \alpha}}{dX_2}\frac{dX_{2}}{dt},
\]
and the emissivity is-
\[\frac{dN_{\sigma \alpha}}{dX_i}=\frac{1}{2\pi}\sum_{\beta} \Im
(\frac{\partial s_{\sigma \alpha \beta}}{\partial X_i} s^{*}_{\sigma
  \alpha \beta})\] with $\alpha=1,3,4$ and the summation over $\beta$
is for all channels $1,2,3,4$. The channels $3$ and $4$ are coupled to
the voltage probe. The coefficient
\[
K_{in,\sigma
  1}=\frac{T_{\sigma,31}+T_{\sigma,41}}{T_{\sigma,31}+T_{\sigma,41}+
  T_{\sigma,32}+T_{\sigma,42}}
\]
multiplied with the second term inside square brackets of Eq.~14,
takes care of the re-injected electrons and hence current
conservation. $T_{\sigma i j}$'s are the transmission coefficients
from lead $j$ to lead $i$.  The above formula is for pumped current
into lead (or, channel) $1$ in presence of inelastic scattering and it
is for the general case. For the special case of very weak pumping one
can analogously as in section III, derive an expression for the pumped
currents as follows-
\begin{figure*}[b]
\protect\centerline{\epsfxsize=7.0in\epsfbox{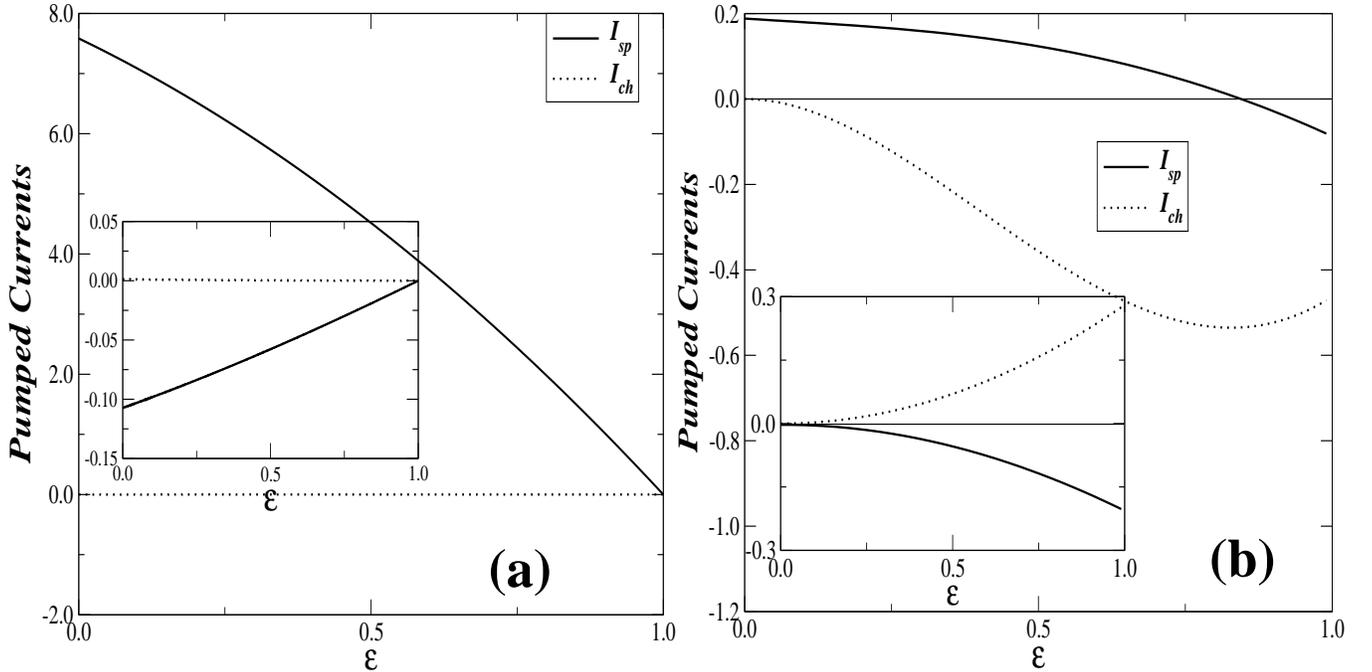}}
\caption{ Dependence of the pumped current on inelastic scattering
  $\epsilon$. Spin polarized pumping delivering a net spin current
  along-with a vanishing charge current in(a) Weak pumping regime. The
  pumped currents are normalized by $I_0$. The parameters are
  $d_{0}=5.0, B_{x}=5.0, \phi=\pi/2, g^{*}=0.44, E/E_{0}=21.56$ and
  wave-vector $q=0$. In the inset the pumped currents are plotted for
  non-resonant case $ E/E_{0}=22.17$ all other parameters remaining
  same. In (b) case of Strong pumping is considered. The parameters
  are $d_{0}=5.0, B_{x}=5.0, x_{p}=1.0, \phi=\pi/2, g^{*}=0.44,
  E/E_{0}=23.0$ and wave-vector $q=0$. In the inset the pumped
  currents are plotted for non-resonant case $ E/E_{0}=38.0$ all other
  parameters remaining same }
\end{figure*}
\begin{eqnarray}
I_{\sigma 1}=\frac{ew x^{2}_{p} sin(\phi)}{2\pi} [J_{\sigma
  1}+K_{in,\sigma 1}(J_{\sigma 3}+J_{\sigma 4})]
\end{eqnarray}

where $K_{in,\sigma 1}$ is as given above while $J_{\sigma f}$'s are
given by-

\begin{eqnarray}
J_{\sigma f}=\sum_{\beta} \Im (\frac{\partial s^{*}_{\sigma f
    \beta}}{\partial X_1}\frac{\partial s^{}_{\sigma f \beta}}{\partial X_2})
\end{eqnarray}

where the summation over $\beta$ runs over all channels $1,2,3$ and
$4$. An unique feature of including inelastic scattering in quantum
pumps is that a new new physical mechanism of rectification comes into
play in the fully incoherent limit. We consider the model system as in
Fig.~7. The model system is coupled to a dephasing reservoir
$\mu_\phi$ via a wave-splitter located at $x=0$. This wave-splitter is
described by the S-Matrix-
\[S_{in}=\left(\begin{array}{cccc}
0    &  \sqrt{1-\epsilon} &  \sqrt{\epsilon} & 0\\
 \sqrt{1-\epsilon}  & 0 &  0 &\sqrt{\epsilon} \\
  \sqrt{\epsilon}     &   0 &  0 & - \sqrt{1-\epsilon}\\
0 &  \sqrt{\epsilon} & - \sqrt{1-\epsilon} & 0 
\end{array} \right) \]

Here the coupling parameter $\epsilon$ characterizes the strength of
inelastic interactions. At $\epsilon=1$, all electrons are
inelastically scattered within the system, whereas at $\epsilon=0$,
the fictitious channels $3\&4$ are effectively decoupled from the
system. In Fig.8(a) and (b), we plot the effect of inelastic
interactions on both the pumped spin and charge currents for incident
energy corresponding to a resonance in the system. In the special case
of very weak pumping the charge current is essentially zero
throughout, while spin-current decreases throughout till the maximum
$\epsilon=1$ is reached but for strong pumping we see that the pumped
charge current increases and in the $\epsilon\rightarrow 1$ limit
dominates the spin current. In the insets of (a) and (b) we plot the
currents for non-resonant pumping, and herein the results in both
cases do not differ much from the resonant case. With inelastic
scattering the device still pumps spin current but now the pumped
charge dominates but only in the strong pumping regime. In the weak
pumping regime the pumped charge current is again zero throughout the
range of the inelastic scattering parameter $\epsilon$. Of-course the
model defined in Ref.\onlinecite{mosk} and utilized earlier in
analyzing inelastic effects in resonant tunneling
diodes\cite{buti_ibm} assumes that an inelastic event takes place only
at a particular point of the whole system. This is depicted in Fig.~7
by the triangle, the junction between the voltage probe and our model
system. A more realistic model would be to couple the system to many
such voltage probes at many points throughout the system. The relevant
parameter in this model is then the probability that the electron be
inelastically scattered while traversing the system. This parameter
can be chosen to be exp($-{d}/{d_\phi}$) as in
Ref.\onlinecite{IBM_gefen}, wherein $d$ is as defined above while
$d_\phi$ is the phase coherence length. This phase coherence length
can be expressed in terms of the dephasing time ($\tau_\phi$), as
$d_{\phi}=v_{f}\tau_{\phi}$ wherein $v_f$ is the Fermi velocity of
electrons traversing the system.  Quantum mechanical coherence is lost
on length scales larger than $d_\phi$. In the 4X4 S-Matrix defined
above to take into account inelastic scattering, the inelastic
scattering parameter $\epsilon$ can be re-parametrized as
1-exp$(-{d}/{d_\phi})$, to obviate this deficiency of single point
inelastic scattering. For complete elastic scattering, i.e., in the
limit $d_{\phi} \gg d, \epsilon\rightarrow0$, while for complete
inelastic scattering, $d_{\phi} \ll d, \epsilon\rightarrow1$.

\section {Noiseless transport}
The adiabatic quantum pump not only generates an electric current but
also heat current which is the sum of the noise and power of joule
heat. A quantum pump is termed optimal if it is noiseless\cite{avron},
i.e., if the total heat generated is only due to the joule heating.
Following the procedure outlined in
Refs.[\onlinecite{wang_heat,wang_opt}], one can derive an elementary
formula for the heat current, joule heat and noise produced in the
pumping mechanism.

The electric current generated in the pumping process and as in Eq.~4
can be reformulated as-

\begin{eqnarray}
I_{\sigma\alpha}=\frac{i e}{2\pi \tau}\int_{0}^{\tau}dt
\sum_{j=1,2}[\partial_{X_{j}}
S_{\sigma}S_{\sigma}^{\dagger}]^{}_{\alpha\alpha}\frac{\partial
  X_{j}}{\partial t}
\end{eqnarray}
The above formula is derived from the more general expression
\begin{eqnarray}
I_{\sigma\alpha}=\frac{e}{\pi \tau}\int_{0}^{\tau}dt\int
dE[S_{\sigma}(E,t)[f(E+i\partial_{t}/2)-f(E)]S_{\sigma}^{\dagger}(E,t)]^{}_{
\alpha\alpha}
\end{eqnarray}

One goes from Eq.~18 to Eq.~17 in the zero temperature limit and by
expanding the Fermi Dirac distribution $f(E+i\partial_{t}/2)$ up-to
first order in $\partial_t$. See for details,
Ref.[\onlinecite{wang_opt}]. One should also keep in mind the fact
that by unitarity of the S-Matrix:
$-i(\partial_{t}S_{\sigma}^{\dagger}S_{\sigma})_{\alpha\alpha}$ equals
$\Im[\partial_{t}S_{\sigma}^{\dagger}S_{\sigma}]_{\alpha\alpha}$. Here
$S_\sigma$ is the 2X2 S-Matrix as defined in section III. Again
$[...]^{ }_{\alpha \alpha}$ represents the $\alpha \alpha ^{th}$
element of the S-Matrix.

The heat current is defined as electric current multiplied by energy
measured from the Fermi level-

\begin{eqnarray}
H_{\sigma\alpha}=\frac{1}{\pi \tau}\int_{0}^{\tau}dt\int
dE(E-E_{F})[S_{\sigma}(E,t)[f(E+i\partial_{t}/2)-f(E)]S_{\sigma}^{\dagger}(E,t)
]^{}_{\alpha\alpha}
\end{eqnarray}
Expanding $f(E+i\partial_{t}/2)$ up-to second order in $\partial_t$,
one gets the heat current in the zero temperature limit as-

\begin{eqnarray}
H_{\sigma\alpha}=\frac{1}{8 \pi
  \tau}\int_{0}^{\tau}dt[\partial_{t}S_{\sigma}(E,t)\partial_{t}S_{\sigma}^{
\dagger}(E,t)]^{}_{\alpha\alpha}
\end{eqnarray}

Here the scattering matrix alluded to- ``$S_\sigma$'', is same as that
in section III, with elements given in Ref.\onlinecite{sm}.

The heat current can be expressed as sum of joule heat and noise as
follows-
\begin{eqnarray}
H_{\sigma\alpha}&=& \frac{1}{8 \pi
  \tau}\int_{0}^{\tau}dt[\partial_{t}S_{\sigma}(E,t)\partial_{t}S_{\sigma}^{
\dagger}(E,t)]^{}_{\alpha\alpha}\nonumber\\
&=&\frac{1}{8 \pi
  \tau}\int_{0}^{\tau}dt[\partial_{t}S_{\sigma}(E,t)S_{\sigma}^{\dagger}(E,t)
S_{\sigma}(E,t)\partial_{t}S_{\sigma}^{\dagger}(E,t)]^{}_{\alpha\alpha}
\nonumber\\
&=&\frac{1}{8 \pi
  \tau}\int_{0}^{\tau}dt\sum_{\beta=1,2}[\partial_{t}S_{\sigma}(E,t)S_{\sigma}^
{\dagger}(E,t)]^{}_{\alpha\beta}[S_{\sigma}(E,t)\partial_{t}S_{\sigma}^{\dagger
}(E,t)]^{}_{\beta\alpha}
\end{eqnarray}

The diagonal term is identified as the joule heat while the
off-diagonal term is the noise\cite{wang_opt}. For $\alpha=1,
\beta=1,2$, the expression for the heat current can be shown to be
broken into the noise and joule parts as follows-
\begin{eqnarray}
H_{\sigma 1}&=&J_{\sigma 1}+N_{\sigma 1}\nonumber\\
&=&\frac{1}{8 \pi \tau}\int_{0}^{\tau}dt[\partial_{t}S_{\sigma}(E,t)S_{\sigma}^{\dagger}(E,t)]^{}_{11}[S_{\sigma}(E,t)\partial_{t}S_{\sigma}^{\dagger}(E,t)]^{}_{11}\nonumber\\
& +&\frac{1}{8 \pi \tau}\int_{0}^{\tau}dt[\partial_{t}S_{\sigma}(E,t)S_{\sigma}^{\dagger}(E,t)]^{}_{12}[S_{\sigma}(E,t)\partial_{t}S_{\sigma}^{\dagger}(E,t)]^{}_{21}
\end{eqnarray}

\begin{figure*}[t]
\protect\centerline{\epsfxsize=7.0in\epsfbox{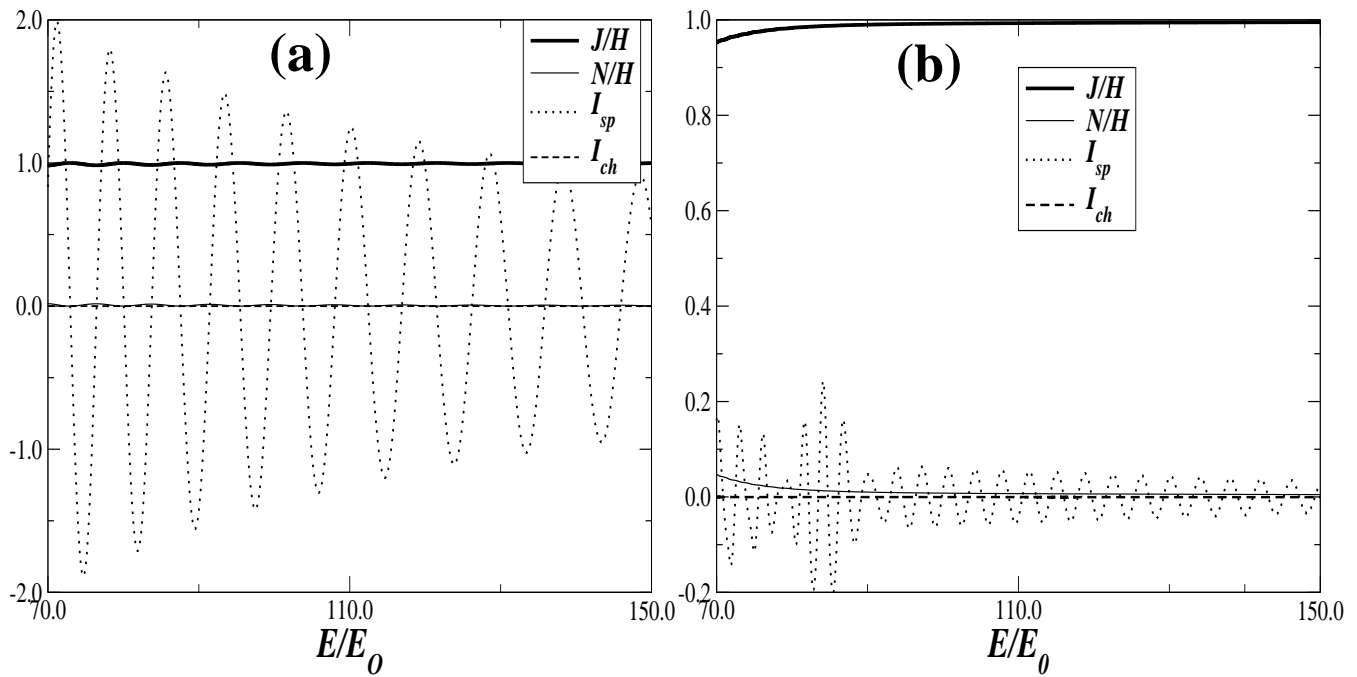}}
\caption{Noise-less transport. (a) The weak pumping regime. Parameters
  are $B_{x}=5.0, d_{0}=5.0, \phi=\pi/2, q=0.0,$ and $g^{*}=0.44$. The
  pumped currents are normalized by $I_0$. (b) The strong pumping
  regime for $x_{p}=6.0, \phi=\pi/10$, all other parameters remaining
  same. }
\end{figure*}

When the pumping amplitude is very small one can similar to previous
cases derive a formula for the heat current, joule heat produced and
noise in our pumping mechanism as has been earlier derived for the
heat current in Ref.\onlinecite{wang_heat} and also for the noise in
Refs.[\onlinecite{mosk_diss,Brou_diss}]. To derive the equations below
we have taken $\tau=2\pi$. Herein below we drop the '$\alpha$' index
in the representation of the heat, joule and noise currents as it is
assumed that we consider currents pumped into lead (or, channel) 1.
Thus-
\begin{eqnarray}
H_{\sigma }&=&\frac{w^2}{16 \pi}[X^{2}_{1}\sum_{\beta=1,2}|\frac{\partial s_{\sigma 1\beta}}{\partial X_{1}}|^{2} +X^{2}_{2}\sum_{\beta=1,2}|\frac{\partial s_{\sigma 1\beta}}{\partial X_{2}}|^{2} + 2 X_{1} X_{2} cos(\phi) \sum_{\beta=1,2} \Re (\frac{\partial s_{\sigma 1\beta}}{\partial X_{1}}  \frac{\partial s^{*}_{\sigma 1\beta}}{\partial X_{2}})]
\end{eqnarray}
\begin{eqnarray}
J_{\sigma }&=&\frac{w^2}{16 \pi}[X^{2}_{1}(\sum_{\beta=1,2}|s^{*}_{\sigma 1\beta}  \frac{\partial s_{\sigma 1\beta}}{\partial X_{1}}|^{2} +2\Re(s_{\sigma 11} s^{*}_{\sigma 12} \frac{\partial s^{*}_{\sigma 11}}{\partial X_1}\frac{\partial s^{ }_{\sigma 12}}{\partial X_1}))
+X^{2}_{2}(\sum_{\beta=1,2}|s_{\sigma 1\beta}\frac{\partial s^{*}_{\sigma 1\beta}}{\partial X_{2}}|^{2} + 2 \Re(s_{\sigma 11} s^{*}_{\sigma 12}\frac{\partial s^{*}_{\sigma 11}}{\partial X_{2}}\frac{\partial s^{ }_{\sigma 12}}{\partial X_{2}}))\nonumber\\
&+& 2 X_{1} X_{2} cos(\phi) ( \sum_{\beta=1,2} |s_{\sigma 1\beta}|^{2}\Re(\frac{\partial s_{\sigma 1\beta}}{\partial X_1}\frac{\partial s^{*}_{\sigma 1\beta}}{\partial X_2})+\Re (s_{\sigma 11} s^{*}_{\sigma 12}\frac{\partial s_{\sigma 12}}{\partial X_{1}} \frac{\partial s^{*}_{\sigma 11}}{\partial X_{2}}) +\Re (s_{\sigma 12} s^{*}_{\sigma 11} \frac{\partial s_{\sigma 11}}{\partial X_{1}}  \frac{\partial s^{*}_{\sigma 12}}{\partial X_{2}}))]
\end{eqnarray}
\begin{eqnarray}
N_{\sigma }&=&\frac{w^2}{16 \pi}[X^{2}_{1}(\sum_{\beta=1,2}|s^{*}_{\sigma 2 \beta} \frac{\partial s^{*}_{\sigma 1\beta}}{\partial X_{1}}|^{2} +2\Re(s_{\sigma 21} s^{*}_{\sigma 22} \frac{\partial s^{*}_{\sigma 11}}{\partial X_1} \frac{\partial s^{ }_{\sigma 12}}{\partial X_1}))+X^{2}_{2}(\sum_{\beta=1,2}|s_{\sigma 2\beta}\frac{\partial s^{*}_{\sigma 1\beta}}{\partial X_{2}}|^{2} +2\Re(s_{\sigma 21}s^{*}_{\sigma 22}\frac{\partial s^{*}_{\sigma 11}}{\partial X_{2}}\frac{\partial s^{ }_{\sigma 21}}{\partial X_{2}}))\nonumber\\
&+&2 X_{1} X_{2} cos(\phi) (\sum_{\beta=1,2} |s_{\sigma 2\beta}|^{2}\Re(\frac{\partial s_{\sigma \beta 1}}{\partial X_2}\frac{\partial s^{*}_{\sigma \beta 1}}{\partial X_1})+\Re (s_{\sigma 21} s^{*}_{\sigma 22}\frac{\partial s_{\sigma 21}}{\partial X_{2}} \frac{\partial s^{*}_{\sigma 11}}{\partial X_{1}}) +\Re (s_{\sigma 22} s^{*}_{\sigma 21} \frac{\partial s_{\sigma 11}}{\partial X_{2}}  \frac{\partial s^{*}_{\sigma 21}}{\partial X_{1}}))]
\end{eqnarray}

In the above equations, '$\Re$' represents the real part of the
complex quantity inside parenthesis.  Since there are no correlations
between electrons with different spin indices\cite{taddei}, the noise
of the charge current and of the spin current is simply
$N=N_{spin}=N_{charge}=N_{+1}+N_{-1}$. Similarly the heat generated
$H=H_{spin}=H_{charge}=H_{+1}+H_{-1}$ and joule heat produced
$J=J_{spin}=J_{charge}=J_{+1}+J_{-1}$.  In Fig.~9, we plot the spin
and charge currents along with the ratio of the power of joule heat to
the heat current ($J/H$) as also the ratio of the noise to heat
current ($N/H$) as function of the Fermi energy. We find that our pump
is completely noiseless throughout the range of Fermi energies in the
weak pumping regime. For very strong pumping in the initial range of
Fermi energies ($E/E_{0}<100.0$) the noise contribution to heat is
small less than $4\%$ while for $E/E_{0}>100.0$ the noise is
negligible less than $0.2\%$ of the total heat generated. Almost all
of the heat generated comes as a result of the joule power. Thus our
model spin pump is almost optimal\cite{avron}. We have also checked
that our spin pump remains optimal for a wide range of variation of
parameters. In the case of weak pumping, the full counting statistics
(distribution of the pumped charge/spin per cycle) is fully
characterized by only two parameters\cite{levitov}, the electric
current generated $I_\sigma$ (Eq.~9) and the noise $N_\sigma$
(Eq.~25). The full counting statistics of our model quantum spin pump
remains as an interesting problem and will be dealt with later on.

\section {Experimental Realization}
To experimentally realize the above proposal, one can apply an
external magnetic field to modulate the strength of magnetization of
the Ferromagnetic stripe and this is one modulating factor, the other
can be apart from the width of the stripe as has been employed in this
work, the distance $Z_{0}$ between stripe and 2DEG, which can be
modulated by applying suitable gate voltages {\it 'a la QPC'}, as has
been done in the first experimental realization of the quantum
pump\cite{switkes_science}. Another method of experimentally realizing
this proposal could be to put two such stripes side by side and
applying different external magnetic fields to both, modulation of
these external fields can effectively provide spin-polarized currents.

\section {Conclusions}

To conclude, a spin polarized device acting on the principles of
quantum adiabatic transport has been proposed. In the dc- transport
case this device does not show any sign of spin-polarization as the
system is time reversal invariant but in the adiabatic regime, when
time reversal invariance is dynamically broken almost cent percent
spin polarization is observed, incidentally in all the simulations,
apart from the effects of inelastic scattering or to a lesser extent
if significant spin-orbit scattering is present, we obtain zero charge
current. As a welcome addition we see almost noiseless transport i.e.,
realization of an optimal quantum spin pump. These features tell us
that a adiabatically modulated magnetic barrier may be the best way to
achieve not only quantum spin pumping but also optimal quantum spin
pumping .


\begin{thebibliography}{99}
\bibitem{sm} The elements of the scattering matrix are given as
   follows:
\begin{eqnarray*}
r_{\sigma}&=&\frac{-i sin(k_{2} d) (k^{2}_{1}-k^{2}_{2}-\lambda^{2}-2i\lambda\sigma k_{1})}{D} \\
t_{\sigma}&=&t^{\prime}_{\sigma}=\frac{2 k_{1} k_{2}}{D}\\
r^{\prime}_{\sigma}&=&\frac{-i sin(k_{2} d) (k^{2}_{1}-k^{2}_{2}-\lambda^{2}+2i\lambda\sigma k_{1})}{D} \\
\mbox{ with }
D&=&2 k_{1} k_{2} cos(k_{2} d) - i sin (k_{2} d)(k^{2}_{1} +k^{2}_{2} +\lambda^{2}), \\
\lambda&=&\frac{g^{*} B_z}{2},\mbox{  }  k_{1}=\sqrt{2E} \mbox{ and } k_{2}=\sqrt{2E-B^{2}_{z}}.
\end{eqnarray*}
It should be noted here that the transmission amplitudes $t_\sigma$
and $t^{\prime}_\sigma$ are spin-independent, spin-polarized pumping
arises because of the spin-dependence of the reflection amplitudes.
\end{thebibliography}
\end{document}